\documentclass[10pt,conference,compsocconf]{IEEEtran} 
\usepackage{threeparttable} 
\usepackage{caption}
\usepackage[noadjust]{cite} 

\ifCLASSINFOpdf
  \usepackage[pdftex]{graphicx}
\else
  \usepackage[dvips]{graphicx}
\fi
\usepackage[cmex10]{amsmath}

\usepackage{eqparbox}
\usepackage{url}

\usepackage{tabularx}
\usepackage{listings}
\usepackage{color}
\definecolor{mygreen}{RGB}{28,172,0} 
\definecolor{mylilas}{RGB}{170,55,241}
\usepackage{xcolor}
\makeatletter
\AtBeginDocument{%
  \let\c@figure\c@lstlisting
  \let\thefigure\thelstlisting
  \let\ftype@lstlisting\ftype@figure 
}
\makeatother

\lstdefinestyle{matlabFile}{
    language=Matlab,
    tabsize=3,
    caption=Test,
    label=code:sample,
    frame=shadowbox,
    rulesepcolor=\color{gray},
    xleftmargin=20pt,
    framexleftmargin=15pt,
    commentstyle=\color{mygreen}, 
    stringstyle=\color{black}, 
    numbers=left,
    numberstyle=\tiny,
    numbersep=5pt,
    breaklines=true,
    showstringspaces=false,
    escapechar=!,
    basicstyle=\ttfamily\footnotesize\bfseries, 
    basewidth  = {.5em,0.4em},
    otherkeywords={set, run, foreach, let, return}, 
    keywordstyle=\color{blue}\bf,
    emph={Test,Message,Implementation, Process_Rank, Num_Processes, avg, standard_deviation, median, max, min, shmem_get, mpi_get, shmem_put, mpi_put, num_pes, rank, mpi_sendrecv, mpi_send_recv, mpi_isend_irecv, Root, HPC_EXP, tr, td, table},emphstyle={\color{magenta}}}
    
\begin{document}
\pagenumbering{gobble}
%
\title{\textbf{\Large Deep Learning:  A Tool for Computational Nuclear Physics}\\
[0.2ex]}

\author{\IEEEauthorblockN{Gianina Alina Negoita\IEEEauthorrefmark{1}\IEEEauthorrefmark{2},
Glenn R. Luecke\IEEEauthorrefmark{3}, 
James P. Vary\IEEEauthorrefmark{4},
Pieter Maris\IEEEauthorrefmark{4},
Andrey M. Shirokov\IEEEauthorrefmark{5}\IEEEauthorrefmark{6}, \\
Ik Jae Shin\IEEEauthorrefmark{7},
Youngman Kim\IEEEauthorrefmark{7},
Esmond G. Ng\IEEEauthorrefmark{8} and
Chao Yang\IEEEauthorrefmark{8}}
\\
\IEEEauthorblockA{\IEEEauthorrefmark{1}Department of Computer Science,
Iowa State University,
Ames, Iowa, USA\\ 
Email: alina@iastate.edu}
\IEEEauthorblockA{\IEEEauthorrefmark{2}Horia Hulubei National Institute for Physics and Nuclear Engineering,
Bucharest-Magurele 76900, Romania}
\IEEEauthorblockA{\IEEEauthorrefmark{3}Department of Mathematics,
Iowa State University,
Ames, Iowa, USA\\ Email: grl@iastate.edu}
\IEEEauthorblockA{\IEEEauthorrefmark{4}Department of Physics and Astronomy,
Iowa State University,
Ames, Iowa, USA\\ Email: jvary@iastate.edu, pmaris@iastate.edu}
\IEEEauthorblockA{\IEEEauthorrefmark{5}Skobeltsyn Institute of Nuclear Physics, Moscow State University, Moscow 119991, Russia\\
Email: shirokov@nucl-th.sinp.msu.ru}
\IEEEauthorblockA{\IEEEauthorrefmark{6}Department of Physics, Pacific National University, Khabarovsk 680035, Russia} 
\IEEEauthorblockA{\IEEEauthorrefmark{7}Rare Isotope Science Project, 
Institute for Basic Science,
Daejeon 34047, Korea\\ Email: geniean@ibs.re.kr, ykim@ibs.re.kr}
\IEEEauthorblockA{\IEEEauthorrefmark{8}Lawrence Berkeley National Laboratory,
Berkeley, California, USA\\ Email: egng@lbl.gov, cyang@lbl.gov}
}


\maketitle

\begin{abstract}
\boldmath
In recent years, several successful applications of the Artificial Neural Networks (ANNs) have emerged in nuclear physics and high-energy physics, as well as in biology, chemistry, meteorology, and other fields of science. 
A major goal of nuclear theory is to predict nuclear structure and nuclear reactions from the underlying theory of the strong interactions, Quantum Chromodynamics (QCD). 
With access to powerful High Performance Computing (HPC) systems, several ab initio approaches, such as the No-Core Shell Model (NCSM), have been developed to calculate the properties of atomic nuclei. However, to accurately solve for the properties of atomic nuclei, one faces immense theoretical and computational challenges. The present study proposes a \textit{feed-forward} ANN method for predicting the properties of atomic nuclei like ground state energy and ground state point proton root-mean-square (rms) radius based on NCSM results in computationally accessible basis spaces. The designed ANNs are sufficient to produce results for these two very different observables in $^6\mathrm{Li}$ from the ab initio NCSM results in small basis spaces that satisfy the theoretical physics condition: independence of basis space parameters in the limit of extremely large matrices. We also provide comparisons of the results from ANNs with established methods of estimating the results in the infinite matrix limit.
\end{abstract}



%
\IEEEpeerreviewmaketitle

\section{Introduction} \label{sec:Introduction}
Nuclei are complicated quantum many-body systems, whose inter-nucleon interactions are not known precisely. The goal of ab initio nuclear theory is to accurately describe nuclei from the first principles as systems of nucleons that interact by fundamental interactions. With sufficiently precise many-body tools, we learn important features of these interactions, such as the fact that three-nucleon (NNN) interactions are critical for understanding the anomalous long lifetime of $^{14}\mathrm{C}$ \cite{PhysRevLett.106.202502}. With access to powerful High Performance Computing (HPC) systems, several ab initio approaches have been developed to study nuclear structure and reactions, such as the No-Core Shell Model (NCSM) \cite{NCSM:2013}, the Green's Function Monte Carlo (GFMC) \cite{MonteCarlo:2001}, the Coupled-Cluster Theory (CC) \cite{CoupledCluster:2004}, the Hyperspherical expansion method \cite{LEIDEMANN:2013158}, the Nuclear Lattice Effective Field Theory \cite{LEE:2009117}\cite{PhysRevLett.106.192501}, the No-Core Shell Model with Continuum \cite{NCSM:2013} and the NCSM-SS-HORSE approach \cite{PhysRevC.94.064320}. These approaches have proven to be successful in reproducing the experimental nuclear spectra for a small fraction of the estimated 7000 nuclei produced in nature.


The ab initio theory may employ a high-quality realistic nucleon-nucleon (NN) interaction, 
which gives an accurate description of NN scattering data and predictions for binding energies, spectra and other observables in light nuclei. Daejeon16 is a NN interaction \cite{SHIROKOV:2016} based on Chiral Effective Field Theory ($\chi$EFT), a promising theoretical approach to obtain a quantitative description of the nuclear force from the first principles \cite{MACHLEIDT:2011}. This interaction has been designed to describe light nuclei without explicit use of NNN interactions, which require a significant increase of computational resources. It has also been shown that this interaction provides good convergence of many-body ab initio NCSM calculations \cite{SHIROKOV:2016}. 

Properties of $^6\mathrm{Li}$ and other nuclei, such as $^3\mathrm{H}$, $^3\mathrm{He}$, $^4\mathrm{He}$, $^6\mathrm{He}$, $^8\mathrm{He}$, $^{10}\mathrm{B}$, $^{12}\mathrm{C}$  and $^{16}\mathrm{O}$, were investigated using the ab initio NCSM approach with the Daejeon16 NN interaction and compared with JISP16 \cite{SHIROKOV:2007} results.
The results showed that Daejeon16 provides both improved convergence and better agreement with data than JISP16.
These calculations were performed with the code MFDn \cite{MFDN:STERNBERG:2008, MFDN:MARIS:2010, MFDN:CPE:2014}, a hybrid MPI/OpenMP code for ab initio nuclear structure calculations. However, one faces major challenges to approach convergence since, as the basis space increases, the demands on computational resources grow very rapidly.

The present work proposes a \textit{feed-forward} Artificial Neural Network (ANN) method as a different approach for obtaining the properties of atomic nuclei such as the ground state (gs) energy and the ground state (gs) point proton root-mean-square (rms) radius based on results from readily-solved basis spaces. \textit{Feed-forward} ANNs can be viewed as universal non-linear function approximators \cite{HORNIK:1989}. Moreover, ANNs can find solution when algorithmic methods are computationally intensive or do not exist. For this reason, ANNs are considered a more powerful modeling method for mapping complex non-linear input-output problems. The output values of ANNs are obtained by simulating the human learning process from the set of learning examples of the input-output association provided to the network. Additional information about ANNs can be found in \cite{BISHOP:ANN:1995}\cite{HAYKIN:ANN:1999}.

Although the gs energy and the gs point proton rms radius are ultimately determined by complicated many-body interactions between the nucleons, the variation of the NCSM calculation results appears to be smooth with respect to the two basis space parameters, $\hbar\Omega$  and $N_{\rm max}$, where $\hbar\Omega$ is the harmonic oscillator (HO) energy and $N_{\rm max}$ is the basis truncation parameter. In practice, these calculations are limited and one can not calculate the gs energy or the gs point proton rms radius for very large $N_{\rm max}$. To obtain the gs energy and the gs point proton rms radius as close as possible to the exact results, the results are extrapolated to the infinite model space. However, it is difficult to construct a simple function with a few parameters to model this type of variation and extrapolate the results to the infinite matrix limit. The advantage of ANN is that it does not need an explicit analytical expression to model the variation of the gs energy or the gs point proton rms radius with respect to $\hbar\Omega$ and $N_{\rm max}$. The \textit{feed-forward} ANN method is very useful to find the converged result at very large $N_{\rm max}$.

In recent years, ANNs have been used in many areas of nuclear physics and high-energy physics. In nuclear physics, ANN models have been developed for constructing a model for the nuclear charge radii \cite{AKKOYUN:2013}, determination of one and two proton separation energies \cite{ProtonSeparationEnergies:ATHANASSOPOULOS:2004}, developing nuclear mass systematics \cite{NuclearMassSystematics:ATHANASSOPOULOS:2004}, identification of impact parameter in heavy-ion collisions \cite{DAVID:1995, BASS:1996, HADDAD:1997}, estimating beta decay half-lives \cite{COSTIRIS:betaDecayHalf-lives:2007} and obtaining potential energy curves \cite{AKKOYUN:potentialEnergyCurves:2013}. In high-energy physics, ANNs are used routinely in experiments for both online triggers and offline data analysis due to an increased complexity of the data and the physics processes investigated. Both the DIRAC \cite{Webpage:DIRAC} and the H1 \cite{Webpage:H1} experiments used ANNs for triggers. For offline data analysis, ANNs were used or tested for a variety of tasks, such as track and vertex reconstruction (DELPHI experiment \cite{FRUHWIRTH:1993}), particle identification and discrimination (decay of the $Z^0$ boson \cite{ABREU:1992}), calorimeter energy estimation and jet tagging. Tevatron experiments used ANNs for the direct measurement of the top quark mass \cite{ABACHI:1997} or leptoquark searches \cite{ABBOTT:1997}.
In terms of types of ANNs, the vast majority of applications in nuclear physics and high-energy physics were based on \textit{feed-forward} ANNs, other types of ANNs remaining almost unexplored. An exception is the DELPHI experiment, which used a recurrent ANN for tracking reconstruction \cite{FRUHWIRTH:1993}.

This research presents results for two very different physical observables for $^6\mathrm{Li}$, gs energy and gs point proton rms radius, produced with the \textit{feed-forward} ANN method. Theoretical data for $^6\mathrm{Li}$ are available from the ab initio NCSM calculations with the MFDn code using the Daejeon16 NN interaction and HO basis spaces up through the cutoff $N_{\rm max} = 18$. This cutoff is defined for $^6\mathrm{Li}$ as the maximum total HO quanta allowed in the Slater determinants forming the basis space less 2 quanta. The dimension of the resulting many-body Hamiltonian matrix is about 2.8 billion at this cutoff. We return to discussing the many-body HO basis shortly. 
However, for the training stage of ANN, data up through $N_{\rm max} = 10$ was used, where the Hamiltonian matrix dimension for $^6\mathrm{Li}$ is only about 9.7 million.
Comparisons of the results from \textit{feed-forward} ANNs with established methods of estimating the results in the infinite matrix limit are also provided.
The paper is organized as follows: In Section~\ref{sec:TheoreticalFramework}, short introductions to the ab initio NCSM method and ANN's formalism are given. In Section~\ref{sec:ANNDesign}, our ANN's architecture is presented. Section~\ref{sec:Results} presents the results and discussions of this work. Section~\ref{sec:Conclusion} contains our conclusion and future work.


\section{Theoretical Framework}\label{sec:TheoreticalFramework}
The NCSM is an ab initio approach to the nuclear many-body problem for light nuclei, which solves for the properties of nuclei for an arbitrary NN interaction, preserving all the symmetries. Naturally, the results obtained with this method are limited to the largest computationally feasible basis space.
We will show that the ANN method is useful to make predictions at ultra-large basis spaces using available data from NCSM calculations at smaller basis spaces. More discussions on these two methods are presented in each subsection.
\subsection{Ab initio NCSM Method}\label{sec:NCSM}
In the NCSM method, the neutrons and protons (separate species of nucleons) interact independently with each other. The Hamiltonian of $A$ nucleons contains kinetic energy ($T_{\rm rel}$) and interaction ($V$) terms
\begin{equation}\label{eq:Hamiltonian}
\begin{split}
	& H_A = T_{\rm rel} + V \\ 
	& = \frac{1}{A} \sum_{i<j} \frac{(\vec{p}_i - \vec{p}_j)^2}{2m} + \sum_{i<j}^{A}V_{ij} + \sum_{i<j<k}^{A}V_{ijk} + ~ \ldots,
\end{split}
\end{equation} 
where $m$ is the nucleon mass,  $\vec{p}_i$ is the momentum of the $i$-th nucleon, $V_{ij}$ is the NN interaction including the Coulomb interaction between protons and $V_{ijk}$ is the NNN interaction. Higher-body interactions are also allowed and signified by the three dots. The HO center-of-mass (CM) Hamiltonian with a Lagrange multiplier is added to the Hamiltonian above to force the many-body eigenstates to factorize into a CM component times an intrinsic component as in \cite{SpuriousCMMotion:1974}. This way, the spurious CM excited states are pushed up above the physically relevant states, which have the lowest eigenstate of the HO for CM motion.

With the nuclear Hamiltonian specified above in~(\ref{eq:Hamiltonian}), the NCSM solves the $A$-body Schr\"{o}dinger equation using a matrix formulation
\begin{equation}\label{eq:Schrodinger}
	H_A \Psi_A(\vec{r}_1, \vec{r}_2, \ldots, \vec{r}_A) = E \Psi_A(\vec{r}_1, \vec{r}_2, \ldots, \vec{r}_A), 
\end{equation}
where the $A$-body wave function is given by a linear combination of Slater determinants $\phi_i$
\begin{equation}\label{eq:Wavefunction}
	\Psi_A(\vec{r}_1, \vec{r}_2, \ldots, \vec{r}_A) = \sum_{i=0}^k c_i \phi_i(\vec{r}_1, \vec{r}_2, \ldots, \vec{r}_A),
\end{equation}
and where $k$ is the number of many-body basis states, configurations, in the system. To obtain the exact A-body wave function one has to consider infinite number of configurations, $k = \infty$. However, in practice, the sum is limited to a finite number of configurations determined by $N_{\rm max}$. The Slater determinant $\phi_i$ is the antisymmetrized product of single particle wave functions $\phi_\alpha(\vec{r})$, where $\alpha$ stands for the quantum numbers of a single particle state. A common choice for the single particle wave functions is the HO basis functions. 
The matrix elements of the Hamiltonian in the many-body HO basis is given by $H_{ij} = \langle \phi_i|\hat{H}|\phi_j \rangle$. For these large and sparse Hamiltonian matrices, the Lanczos method is one possible choice to find the extreme eigenvalues \cite{PARLETT:SymmetricEigenvalueProblem:1998}.

To be more specific, our limited many-body HO basis is characterized by two basis space parameters: $\hbar\Omega$ and $N_{\rm max}$, where $\hbar\Omega$ is the HO energy and $N_{\rm max}$ is the basis truncation parameter. In this approach, all possible configurations with $N_{\rm max}$ excitations above the unperturbed gs (the HO configuration with the minimum HO energy defined to be the $N_{\rm max} = 0$ configuration) are considered.
Even values of $N_{\rm max}$ correspond to states with the same parity as the unperturbed gs and are called the ``natural" parity states, while odd values of $N_{\rm max}$ correspond to states with ``unnatural" parity. 

Due to the strong short-range correlations of nucleons in a nucleus, a large basis space, or model space, one that is often not feasible, is required to achieve convergence. To obtain the gs energy and other observables as close as possible to the exact results one has to choose the largest feasible basis spaces. Next, if numerical convergence is not achieved, which is often the case, the results are extrapolated to the infinite model space. To take the infinite matrix limit, several extrapolation methods have been developed (see, for example, \cite{ExtrapolationMethods:2009}).

\subsection{Artificial Neural Networks}\label{sec:ANNs} 
ANNs are powerful tools that can be used for function approximation, classification and pattern recognition, such as finding clusters or regularities in the data. The goal of ANNs is to find a solution efficiently when algorithmic methods are computationally intensive or do not exist. 
An important advantage of ANNs is the ability to detect complex non-linear input-output relationships. For this reason, ANNs can be viewed as universal non-linear function approximators \cite{HORNIK:1989}. Employing ANNs for mapping complex non-linear input-output problems offers a significant advantage over conventional techniques, such as regression techniques, because ANNs do not require explicit mathematical functions.

ANNs are defined as computer algorithms that mimic the human brain, being inspired by biological neural systems. Similar to the human brain, ANNs can perform complex tasks, such as learning, memorization and generalization. They are capable of learning from experience, storing knowledge and then applying this knowledge to make predictions. 

A biological neuron has a cell body, a nucleus, dendrites and an axon. Dendrites act as inputs, the axon propagates the signal and the interaction between neurons takes place at synapses. Each synapse has an associated weight. When a neuron `fires', it sends an output through the axon and the synapse to another neuron. Each neuron then collects all the inputs coming from linked neurons and produces an output.

\begin{figure}[htbp]
\centering%
\includegraphics[width=\linewidth]{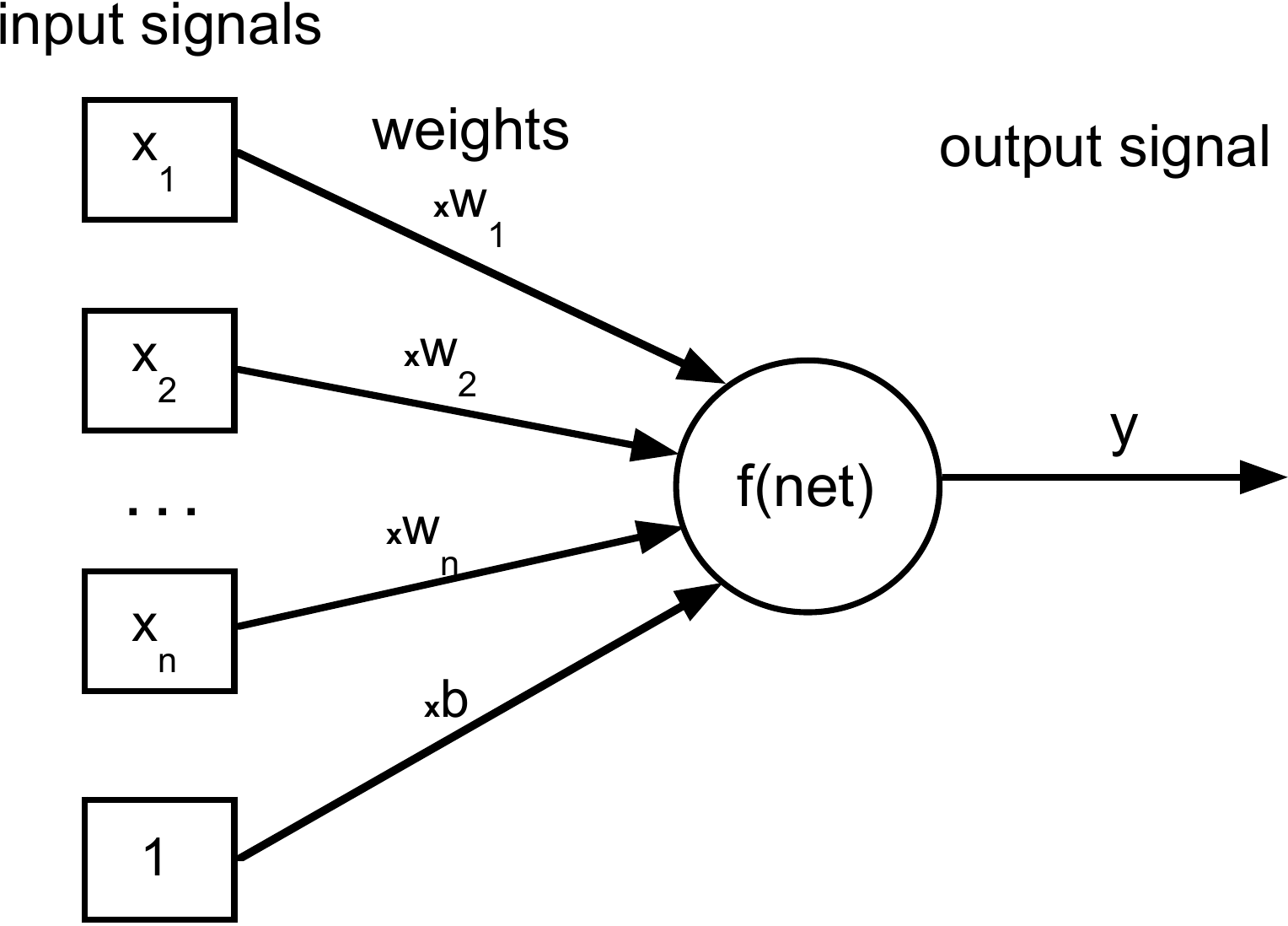} 
\caption{An artificial neuron.}\label{fig:ArtificialNeuron}
\end{figure}

The artificial neuron (AN) is a model of the biological neuron. Figure~\ref{fig:ArtificialNeuron} shows a representation of an AN. Similarly, the AN receives a set of input signals $(x_1, x_2, ..., x_n)$ from an external source or from another AN. A weight $w_i$ $(i = 1, ..., n)$ is associated with each input signal $x_i$ $(i = 1, ..., n)$. Additionally, each AN that is not in the input layer has another input signal called the \textit{bias} with value 1 and its associated weight $b$. The AN collects all the input signals and calculates a \textit{net} signal as the weighted sum of all input signals as

\begin{equation}\label{eq:net}
	net = \sum_{i=1}^{n+1} w_i x_i,
\end{equation}
where $x_{n+1} = 1$ and $w_{n+1} = b$. 

Next, the AN calculates and transmits an output signal, $y$. The output signal is calculated using a function called an \textit{activation} or \textit{transfer} function, which depends on the value of the \textit{net} signal, $y = f(net)$.

ANNs consist of a number of highly interconnected ANs which are processing units. One simple way to organize ANs is in layers, which gives a class of ANN called multi-layer ANN. ANNs are composed of an input layer, one or more hidden layers and an output layer. The neurons in the input layer receive the data from outside and transmit the data via weighted connections to the neurons in the hidden layer, which, in turn, transmit the data to the next layer. Each layer transmits the data to the next layer. Finally, the neurons in the output layer give the results. The type of ANN, which propagates the input through all the layers and has no \textit{feed-back} loops is called a \textit{feed-forward} multi-layer ANN. For simplicity, throughout this paper we adopt and work with a \textit{feed-forward} ANN. For other types of ANN, see \cite{BISHOP:ANN:1995}\cite{HAYKIN:ANN:1999}.

\begin{figure}[htbp]
\centering%
\includegraphics[width=\linewidth]{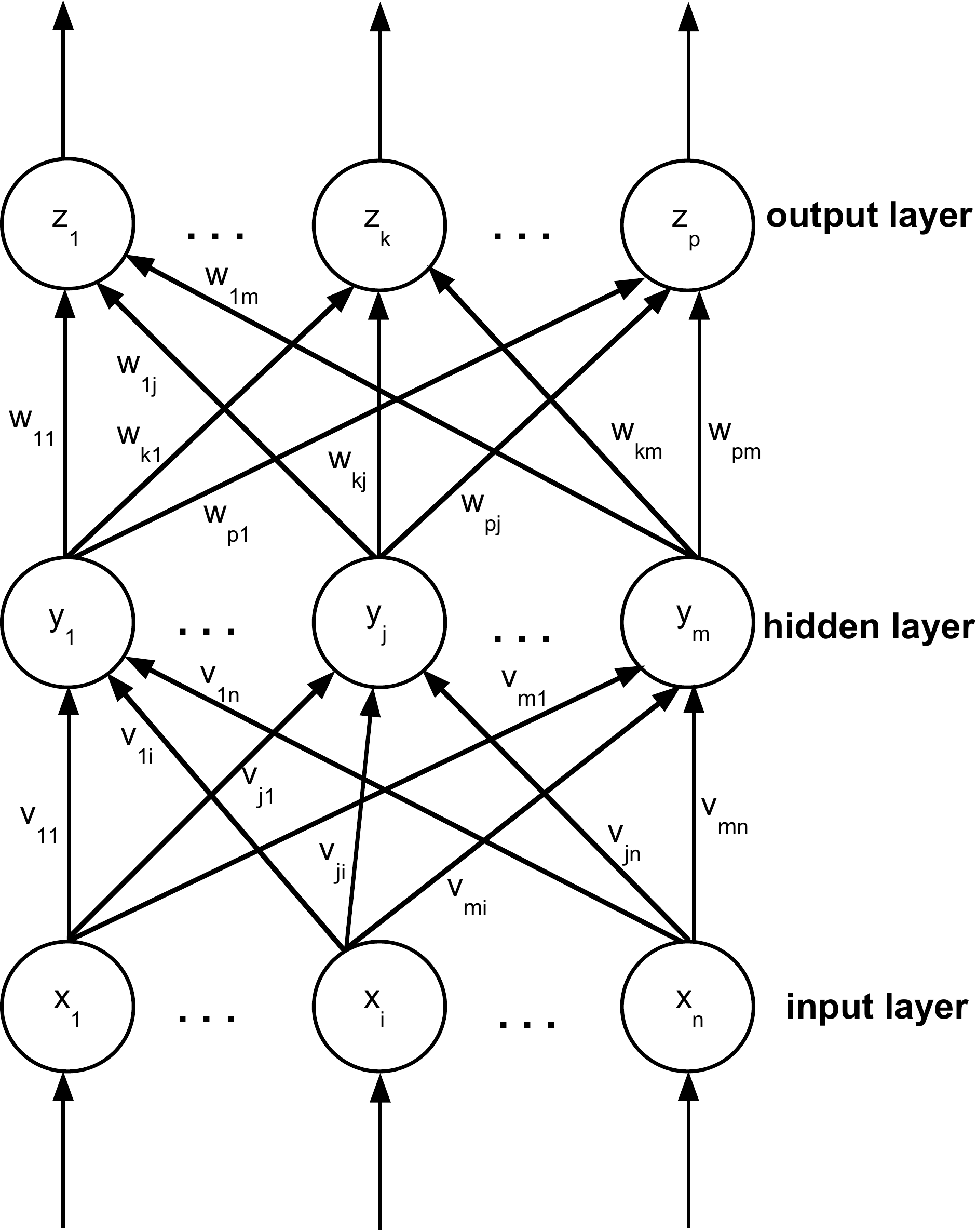} 
\caption{A three-layer ANN.}\label{fig:ANN}
\end{figure}

Figure~\ref{fig:ANN} shows an example of a \textit{feed-forward} three-layer ANN. It contains one input layer, one hidden layer and one output layer. The input layer has $n$ ANs, the hidden layer has $m$ ANs and the output layer has $p$ ANs. The connections between the neurons are weighted as follows: $v_{ji}$ are the weights between the input layer and the hidden layer, and $w_{kj}$ are the weights between the hidden layer and the output layer, where ($i = 1, ..., n$), ($j = 1, ..., m$) and ($k = 1, ..., p$). In this example, the input layer has no \textit{activation} function, the hidden layer has \textit{activation} function $f$ and the output layer has \textit{activation} function $g$. It is also possible to have a different \textit{activation} function for each individual neuron.

The \textit{activation} function in the hidden layer, $f$, is different from the \textit{activation} function in the output layer, $g$. For function approximation, a common choice for the \textit{activation} function for the neurons in the hidden layer is a \textit{sigmoid} or \textit{sigmoid}--like function, while the neurons in the output layer have a \textit{linear} function: 
\begin{equation}
	f(x) = \frac {1} {1 + e^{-ax}},
\end{equation}
where $a$ is the slope parameter of the \textit{sigmoid} function and
\begin{equation}
	g(x) = x.
\end{equation}
The neurons with \textit{non-linear} \textit{activation} functions allow the ANN to learn \textit{non-linear} and \textit{linear} relationships between input and output vectors. Therefore, sufficient neurons should be used in the hidden layer in order to get a good function approximation. 

In the example shown in Figure~\ref{fig:ANN} and with the notations mentioned above, the network propagates the external signal through the layers producing the output signal $z_k$ at neuron $k$ in the output layer

\begin{equation}\label{eq:output}
\begin{split}
	z_k = g(net_{z_{k}}) & = g(\sum_{j=1}^{m+1} w_{kj} f(net_{y_{j}})) \\
	& = g(\sum_{j=1}^{m+1} w_{kj} f(\sum_{i=1}^{n+1} v_{ji} x_i)).
\end{split}
\end{equation}

The use of an ANN is a two-step process, training and testing stages. In the training stage, the ANN adjusts its weights until an acceptable error level between desired and predicted outputs is obtained. The difference between desired and predicted outputs is measured by the error function, also called the performance function. A common choice for the error function is \textit{mean square error} (MSE).

There are multiple training algorithms based on various implementations of the \textit{back-propagation} algorithm \cite{HAGAN:backpropagation:1994}, an efficient method for computing the gradient of error functions. These algorithms compute the \textit{net} signals and outputs of each neuron in the network every time the weights are adjusted as in (\ref{eq:output}), the operation being called the \textit{forward pass} operation. Next, in the \textit{backward pass} operation, the errors for each neuron in the network are computed and the weights of the network are updated as a function of the errors until the stopping criterion is satisfied. 
In the testing stage, the trained ANN is tested over new data that was not used in the training process. The predicted output is calculated using (\ref{eq:output}). 

One of the known problems for ANN is overfitting: the error on the training set is within the acceptable limits, but when new data is presented to the network the error is large. In this case, ANN has memorized the training examples, but it has not learned to generalize to new data. This problem can be prevented using several techniques, such as early stopping, regularization, weight decay, \textit{hold-out} method, \textit{m-fold cross-validation} and others. 

Early stopping is widely used. In this technique the available data is divided into three subsets: the training set, the validation set and the test set. The training set is used for computing the gradient and updating the network weights and biases. The error on the validation set is monitored during the training process. When the validation error increases for a specified number of iterations, the training is stopped, and the weights and biases at the minimum of the validation error are returned. The test set error is not used during training, but it is used as a further check that the network generalizes well and to compare different ANN models.

Regularization modifies the performance function by adding a term that consists of the mean of the sum of squares of the network weights and biases. However, the problem with regularization is that it is difficult to determine the optimum value for the performance ratio parameter. It is desirable to determine the optimal regularization parameters automatically. One approach to this process is the Bayesian regularization of David MacKay \cite{MacKay:bayesianinterpolation:1992}. The Bayesian regularization algorithm updates the weight and bias values according to \textit{Levenberg-Marquardt} \cite{HAGAN:backpropagation:1994}\cite{Marquardt:LM:1963} optimization. It minimizes a linear combination of squared errors and weights and it also modifies the regularization parameters of the linear combination to generate a network that generalizes well. See \cite{MacKay:bayesianinterpolation:1992}\cite{FORESEE:1997} for more detailed discussions of Bayesian regularization.

For further and general background on the ANN and how to prevent overfitting and improve generalization refer to \cite{BISHOP:ANN:1995}\cite{HAYKIN:ANN:1999}.


\section{ANN Design} \label{sec:ANNDesign}
\begin{figure}[htbp]
\centering%
\includegraphics[width=\linewidth]{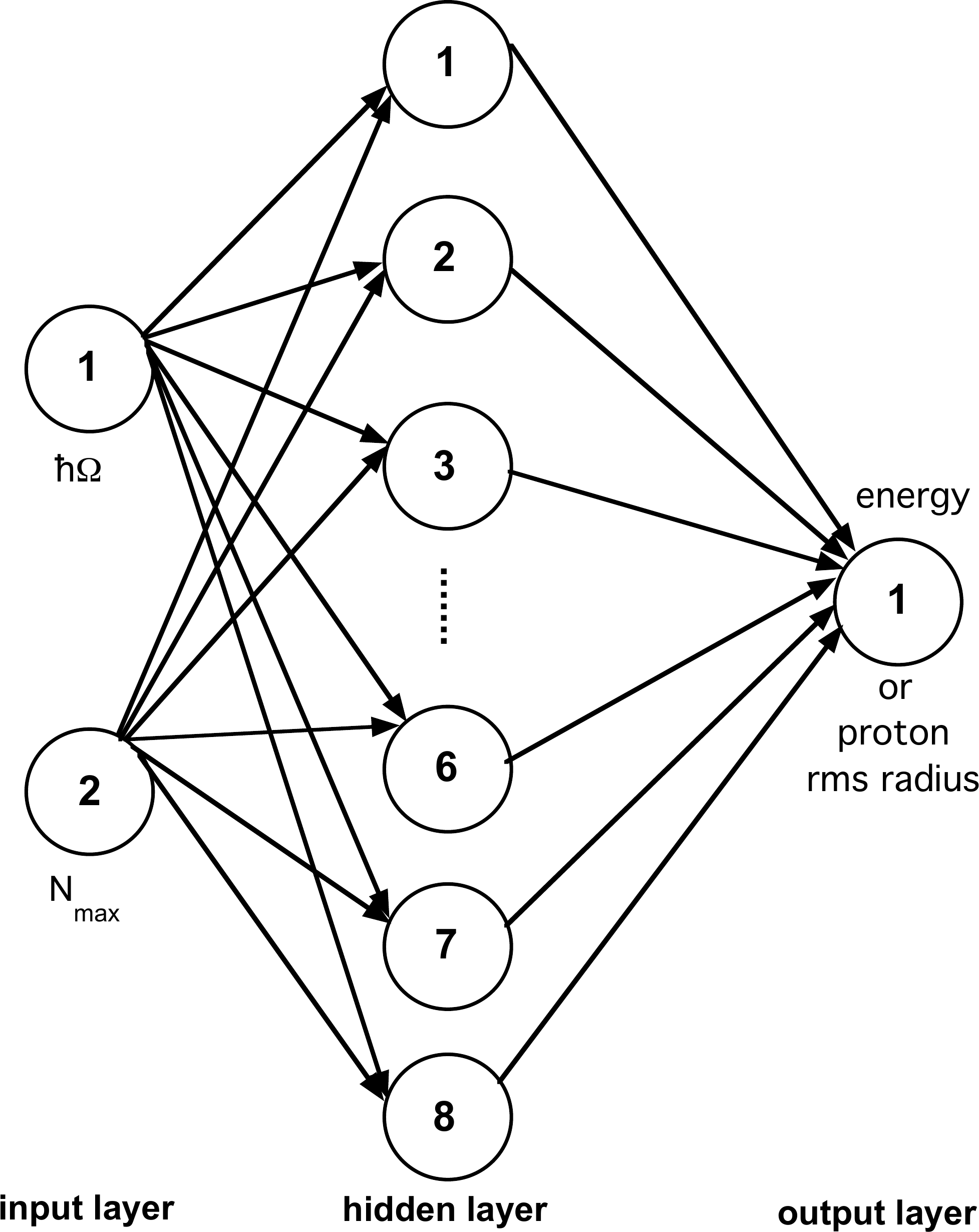} 
\caption{Topological structure of the designed ANN.}\label{fig:ANN_architecture}
\end{figure}

The topological structure of ANNs used in this study is presented in Figure~\ref{fig:ANN_architecture}. The designed ANNs contain one input layer with two neurons, one hidden layer with eight neurons and one output layer with one neuron. The inputs were the basis space parameters: the HO energy, $\hbar\Omega$, and the basis truncation parameter, $N_{\rm max}$, described in Section~\ref{sec:TheoreticalFramework}. The desired outputs were the gs energy and the gs point proton rms radius of $^6\mathrm{Li}$. An ANN was designed for each desired output: one ANN for gs energy and another ANN for gs point proton rms radius. The optimum number of neurons in the hidden layer was obtained according to a trial and error process. 

The \textit{activation} function employed for the hidden layer was a widely-used form, the \textit{hyperbolic tangent sigmoid} function
\begin{equation}\label{eq:tansig}
	f(x) = tansig(x) = \frac{2}{(1 + e^{-2x})} - 1, 
\end{equation}
where $x$ is the input value of the hidden neuron and $f(x)$ is the output of the hidden neuron. $tansig$ is mathematically equivalent to the \textit{hyperbolic tangent function}, $tanh$, but it improves network functionality because it runs faster than $tanh$.
It has been proven that one hidden layer and \textit{sigmoid}-like \textit{activation} function in this layer are sufficient to approximate any continuous real function, given sufficient number of neurons in the hidden layer \cite{CYBENKO:1989}.

MATLAB software v9.2.0 (R2017a) with \textit{Neural Network Toolbox} was used for the implementation of this work. As mentioned before in Section~\ref{sec:Introduction}, the data set for $^6\mathrm{Li}$ was taken from the ab initio NCSM calculations with the MFDn code using the Daejeon16 NN interaction \cite{SHIROKOV:2016} and basis spaces up through $N_{\rm max} = 18$. However, only the data with even $N_{\rm max}$ values corresponding to ``natural" parity states and up through $N_{\rm max} = 10$ was used for the training stage of the ANN. The training data was limited to $N_{\rm max} = 10$ and below since future applications to heavier nuclei will likely not have data at higher $N_{\rm max}$ values due to exponential increase in the matrix dimension.
This $N_{\rm max} \leq 10$ data set was randomly divided into two separate sets using the \textit{dividerand} function in MATLAB: 85\% for the training set and 15\% for the testing set. 
A \textit{back-propagation} algorithm with Bayesian regularization with MSE performance function was used for ANN training. Bayesian regularization does not require a validation data set. 

\begin{figure}[htbp]
\centering%
\includegraphics[width=\linewidth]{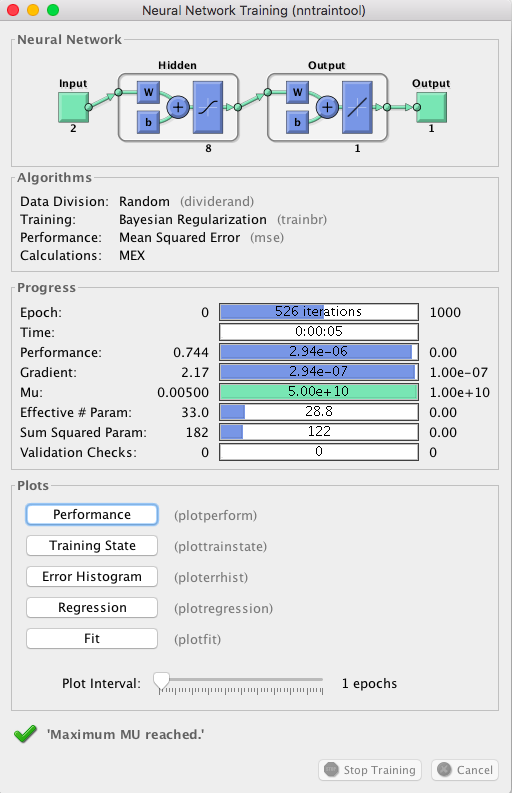} 
\caption{Neural Network Training tool (nntraintool) in MATLAB.}\label{fig:nntraintool}
\end{figure}

For function approximation, Bayesian regularization provides better generalization performance than early stopping in most cases, but it takes longer to converge. The performance improvement is more noticeable when the data set is small because Bayesian regularization does not require a validation data set, leaving more data for training. 
In MATLAB, Bayesian regularization has been implemented in the function \textit{trainbr}. When using \textit{trainbr}, it is important to train the network until it reaches convergence.
In this study, the training process is stopped if: (1) it reaches the maximum number of iterations, 1000; (2) the performance has an acceptable level; (3) the estimation error is below the target; or (4) the Levenberg-Marquardt adjustment parameter $\mu$ becomes larger than $10^{10}$. A good typical indication for convergence is when the maximum value of $\mu$ has been reached.
During training, one can choose to show the Neural Network Training tool (nntraintool) GUI in MATLAB to monitor the training progress. Figure~\ref{fig:nntraintool} illustrates a training example as it appears in nntraintool.

Note the ANN architecture view and the training stopping parameters with their ranges. 

\section{Results and Discussions} \label{sec:Results}
Every ANN creation and initialization function starts with different initial conditions, such as initial weights and biases, and different division of the training, validation, and test data sets. These different initial conditions can lead to very different solutions for the same problem. Moreover, it is also possible to fail in obtaining realistic solutions with ANNs for certain initial conditions. For this reason, it is a good idea to train several networks to ensure that a network with good generalization is found. Furthermore, by retraining each network, one can verify a robust network performance. 

Figure~\ref{trainingExample} shows the training procedure of 100 ANNs with architecture mentioned in Section~\ref{sec:ANNDesign} using the \textit{trainbr} function for Bayesian regularization. 
Each ANN is trained starting from different initial weights and biases, and with different division for the training and test data sets. To ensure good generalization, each ANN is retrained 5 times. 
\begin{center}
\begin{minipage}[c]{\linewidth}
\begin{lstlisting}[caption=Training 100 ANNs and retraining each ANN 5 times to find the best generalization., captionpos=b, label={trainingExample}]
net = fitnet(8, 'trainbr');
net.performFcn = 'mse'; 
numNN = 100; 
numNNr = 5;
NN = cell(numNNr, numNN);
trace = cell(numNNr, numNN);
perfs = zeros(numNNr, numNN);
% train numNN ANNs
for i = 1:numNN
   % retrain each ANN numNNr times
   for j = 1:numNNr
     [NN{j}{i},trace{j}{i}] = train(net, x, t); 
     y2 = NN{j}{i}(x2);
     perfs(j, i) = perform(NN{j}{i}, t2, y2);
     net = NN{j}{i};
   end
   % reinitialize initial weights and biases 
   net = init(net);
end
minPerf = min(perfs(:))
[rowMin, colMin] = find(perfs == minPerf)
net = NN{rowMin}{colMin};
tr = trace{rowMin}{colMin};
\end{lstlisting}
\end{minipage}
\end{center}

The performance function, such as MSE, measures how well ANN can predict data, i.e., how well ANN can be generalized to new data. 
The test data sets are a good measure of generalization for ANNs since they are not used in training. A small performance function on the test data set indicates an ANN with good performance was found. In this work, the ANN with the lowest performance on the test data set is chosen to make future predictions.

Using the methodology described above, two ANNs are chosen to predict the gs energy and the gs point proton rms radius. The ANN prediction results for the gs energies and gs proton rms radii of $^6\mathrm{Li}$ are presented in detail in this section. Comparison with the ab initio NCSM calculation results is also provided for the available data at $N_{\rm max} = 12-18$.

\begin{figure}[htbp]
\centering%
\includegraphics[width=\linewidth]{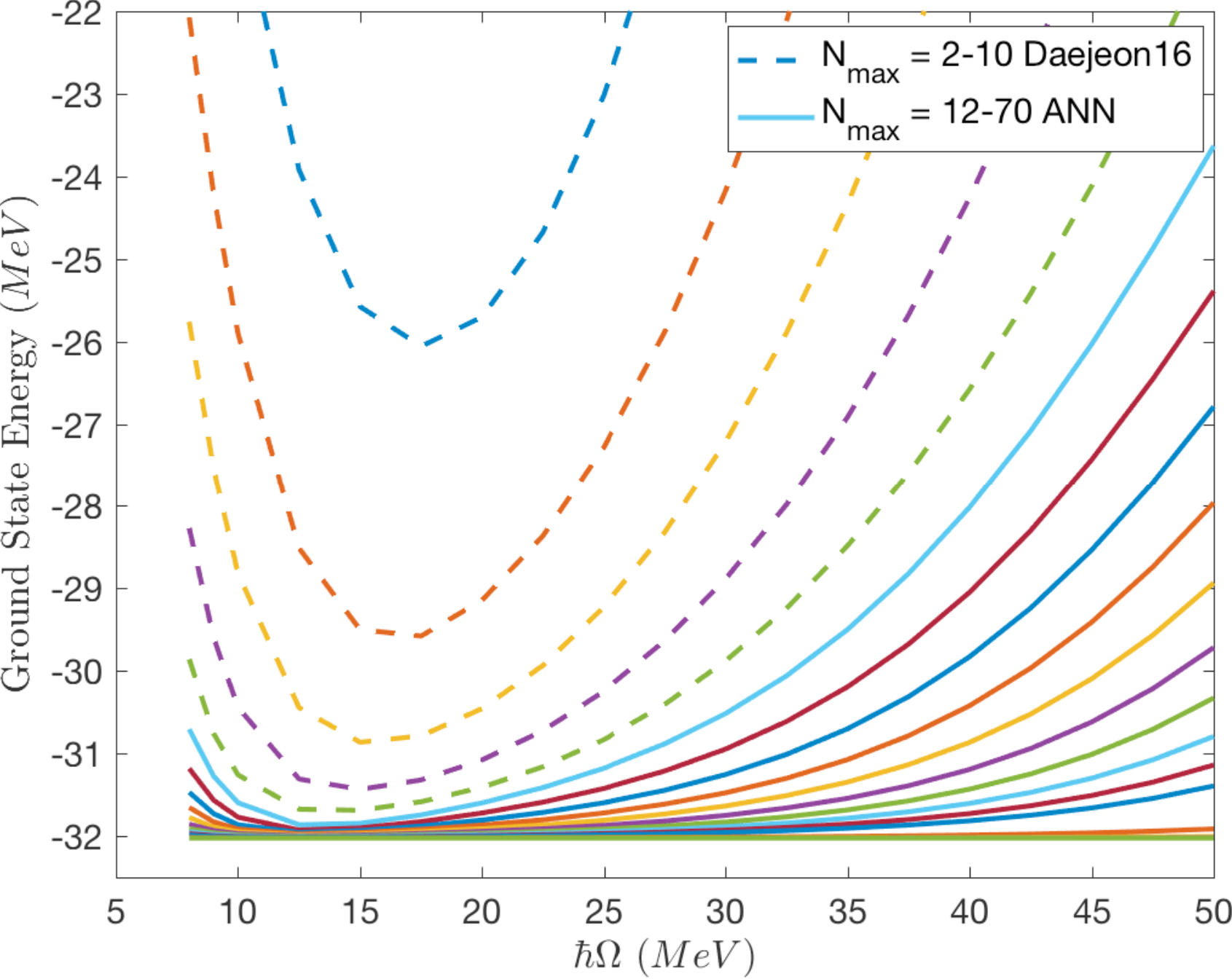} 
\caption{Calculated and predicted gs energy of $^6\mathrm{Li}$ as a function of $\hbar\Omega$ at selected $N_{\rm max}$ values. 
}
\label{fig:Li6_gs_energy}
\end{figure}

Figure~\ref{fig:Li6_gs_energy} presents the gs energy of $^6\mathrm{Li}$ as a function of the HO energy, $\hbar\Omega$, at selected values of the basis truncation parameter, $N_{\rm max}$. The dashed curves connect the NCSM calculation results using the Daejeon16 NN interaction for $N_{\rm max} = 2-10$, in increments of 2 units, used for ANN training and testing. The solid curves link the ANN prediction results for $N_{\rm max} = 12-70$. The sequence from $N_{\rm max} = 12-30$ is in increments of 2 units, while the sequence from $N_{\rm max} = 30-70$ is in increments of 10 units. The lowest horizontal line corresponds to $N_{\rm max} = 70$ and represents the nearly converged result predicted by ANN. Convergence is defined as independence of both basis space parameters, $\hbar\Omega$ and $N_{\rm max}$.
The convergence pattern shows a reduction in the spacing between successive curves and flattening of the curves as $N_{\rm max}$ increases.
The gs energy provided by the ANN decreases monotonically with increasing $N_{\rm max}$ at all values of $\hbar\Omega$. This demonstrates that the ANN is successfully simulating what is expected from theoretical physics. That is, in theoretical physics the energy variational principle requires that the gs energy behaves as a non-increasing function of increasing matrix dimensionality at fixed
$\hbar\Omega$ and, furthermore, matrix dimension increases with increasing $N_{\rm max}$.

\begin{figure}[htbp]
\centering%
\includegraphics[width=\linewidth]{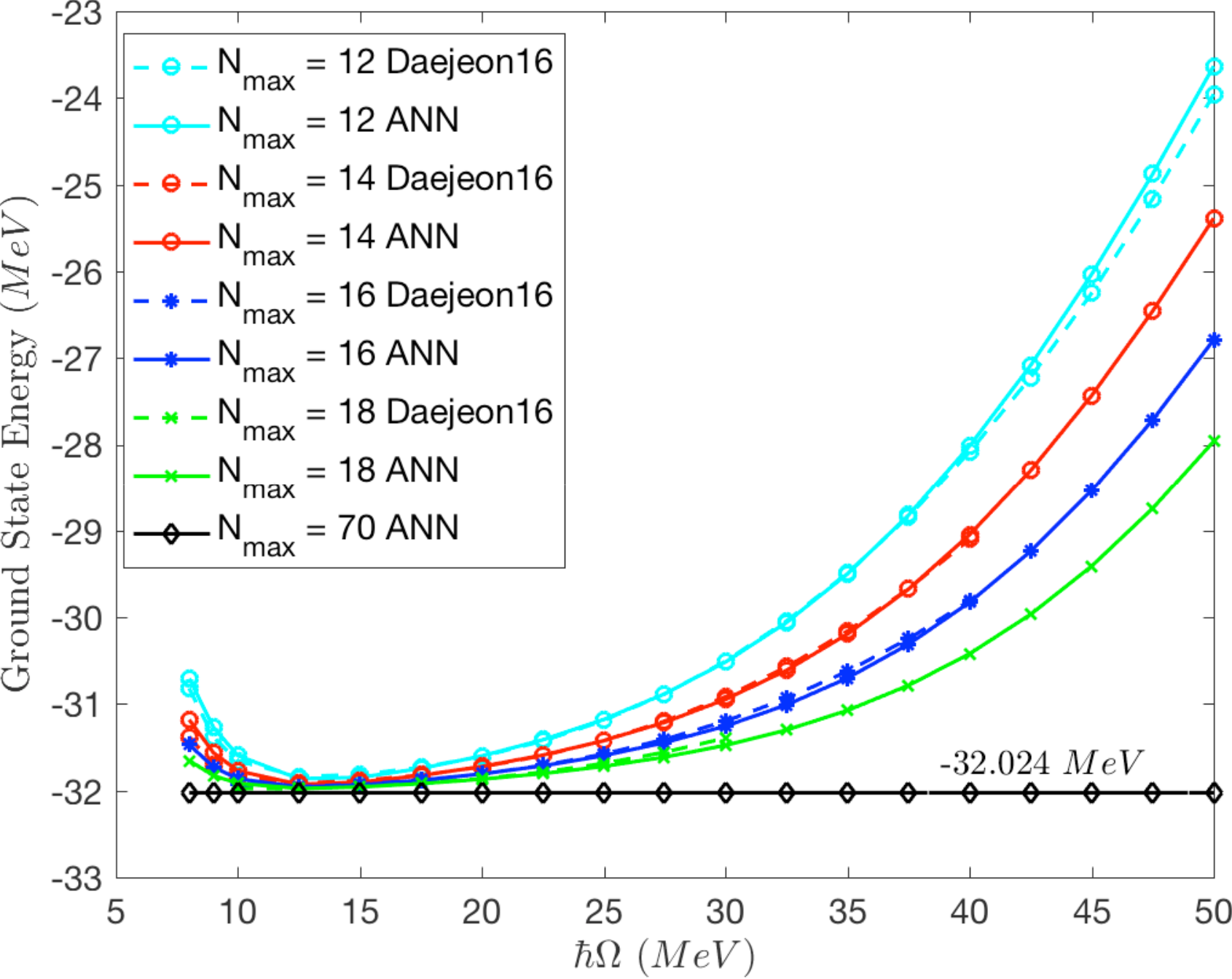} 
\caption{Comparison of the NCSM calculated and the corresponding ANN predicted gs energy values of $^6\mathrm{Li}$ as a function of $\hbar\Omega$ at $N_{\rm max} = 12, 14, 16,$ and $18$. 
The lowest horizontal line corresponds to the ANN nearly converged result at $N_{\rm max} = 70$.}\label{fig:Li6_gs_energy_cp}
\end{figure}

To illustrate the ANN prediction accuracy, the NCSM calculation results and the corresponding ANN prediction results of the gs energy of $^6\mathrm{Li}$ are presented in Figure~\ref{fig:Li6_gs_energy_cp} as a function of $\hbar\Omega$ at $N_{\rm max} = 12, 14, 16$, and $18$. The dashed curves connect the NCSM calculation results using the Daejeon16 NN interaction and the solid curves link the ANN prediction results. The nearly converged result predicted by ANN is also shown above the horizontal axis at $N_{\rm max} = 70$. 
Figure~\ref{fig:Li6_gs_energy_cp} shows good agreement between the calculated NCSM results and the ANN predictions up through $N_{\rm max} = 18$.  Actual NCSM results always converged from above towards the exact result and become increasingly independent of the basis space parameters, $\hbar\Omega$ and $N_{\rm max}$. That the ANN result is essentially a flat line at $N_{\rm max} = 70$ and that the curves preceding it form an increasingly dense pattern approaching $N_{\rm max} = 70$ both provide indications that the ANN is producing a valid estimate of the converged gs energy.

\begin{figure}[htbp]
\centering%
\includegraphics[width=\linewidth]{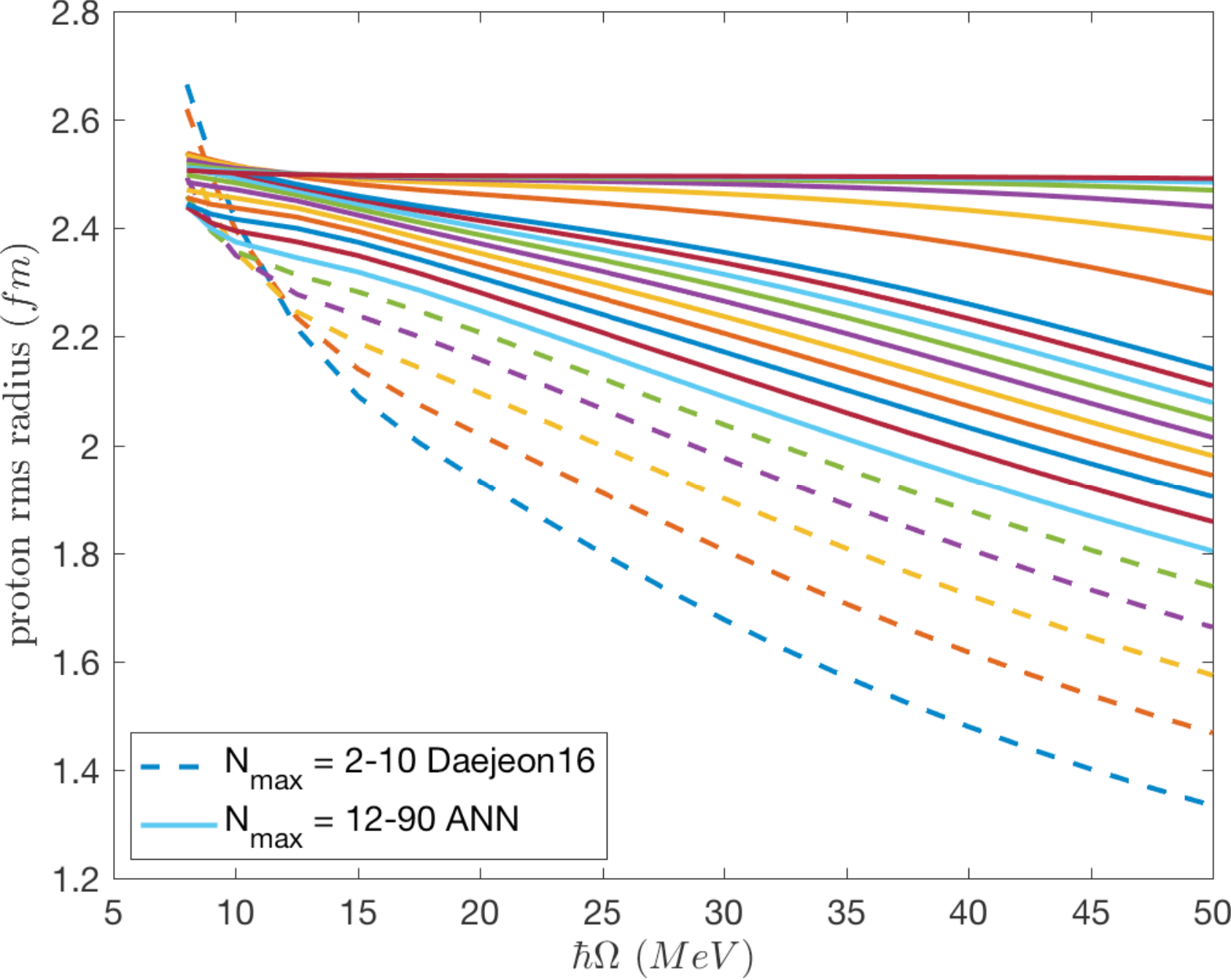} 
\caption{Calculated and predicted gs point proton rms radius of $^6\mathrm{Li}$ as a function of $\hbar\Omega$ at selected $N_{\rm max}$ values. 
}\label{fig:Li6_gs_radius}
\end{figure}

The gs rms radii provide a very different quantity from NCSM results as they are found to be more slowly convergent than the gs energies and they are not monotonic.
Figure~\ref{fig:Li6_gs_radius} presents the calculated gs point proton rms radius of $^6\mathrm{Li}$ as a function of $\hbar\Omega$ at selected values of $N_{\rm max}$. The dashed curves connect the NCSM calculation results using the Daejeon16 NN interaction up through $N_{\rm max}=10$, while the solid curves link the ANN prediction results above $N_{\rm max}=10$. The highest curve corresponds to $N_{\rm max} = 90$ and successively lower curves are obtained with $N_{\rm max}$ decreased by 10 units until the $N_{\rm max}=30$ curve and then by 2 units for each lower $N_{\rm max}$ curve. The rms radius converges monotonically from below for most of the $\hbar\Omega$ range shown. More importantly, the rms radius shows the anticipated convergence to a flat line accompanied by an increasing density of lines with increasing $N_{\rm max}$.  These are the signals of convergence that we anticipate based on experience in limited basis spaces and on general theoretical physics grounds.

\begin{figure}[htbp]
\centering%
\includegraphics[width=\linewidth]{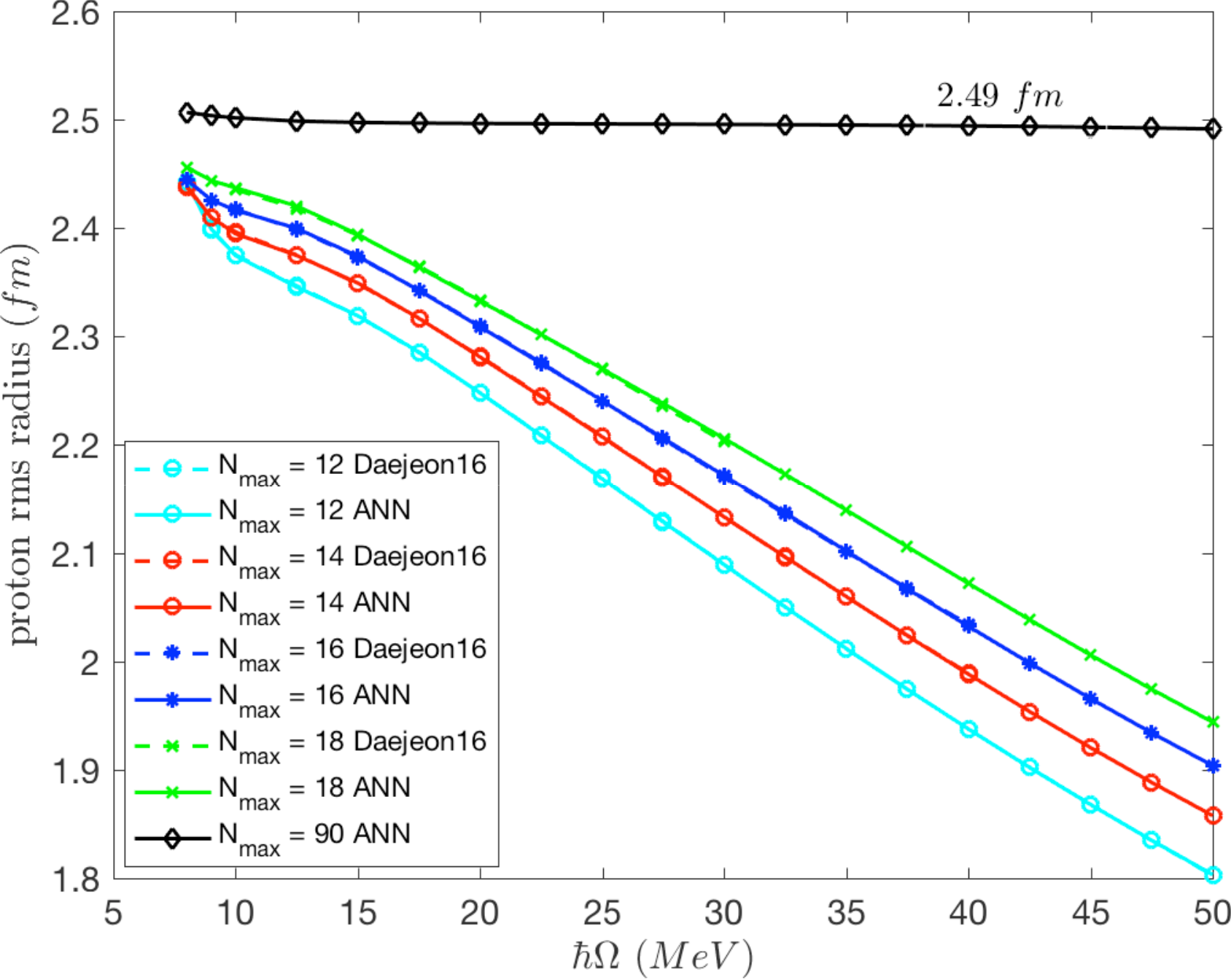} 
\caption{Comparison of the NCSM calculated and the corresponding ANN predicted gs point proton rms radius values of $^6\mathrm{Li}$ as a function of $\hbar\Omega$ for $N_{\rm max} = 12, 14, 16,$ and $18$. 
The highest curve corresponds to the ANN nearly converged result at $N_{\rm max} = 90$.}
\label{fig:Li6_gs_radius_cp}
\end{figure}

The NCSM calculated values and the corresponding prediction values of the gs point proton rms radius of $^6\mathrm{Li}$ are presented in Figure~\ref{fig:Li6_gs_radius_cp} for $N_{\rm max} = 12, 14, 16,$ and $18$. The dashed curves link the NCSM calculation results using the Daejeon16 NN interaction and the solid curves connect the ANN prediction results. As seen in this figure, the ANN predictions are in good agreement with the NCSM calculations, showing the efficacy of the ANN method.

Table~\ref{tab:results} presents the nearly converged ANN predicted results for the gs energy and the gs point proton rms radius of $^6\mathrm{Li}$. As a comparison, the gs energy results from the current best theoretical upper bounds at $N_{\rm max} = 10$ and $N_{\rm max} = 18$ and from the Extrapolation B (Extrap B) method \cite{ExtrapolationMethods:2009} at $N_{\rm max} \leq 10$ are provided. Similar to the ANN prediction, the Extrap B result arises when using all available results through $N_{\rm max}=10$. The ANN prediction for the gs energy is below the best upper bound, found at $N_{\rm max}=18$, which is about 85 $KeV$ lower than the Extrap B result.

There is no extrapolation available for the rms radius, but we quote in Table~\ref{tab:results} the estimated result by the \textit{crossover-point} method \cite{BOGNER:200821} to be $\sim2.40$ $fm$. The \textit{crossover-point} method takes the value at $\hbar\Omega$ in the table of rms radii results through $N_{\rm max}=10$, which produces an rms radius result that is roughly independent of $N_{\rm max}$.


\captionsetup{font={footnotesize,sc},justification=centering,labelsep=period}%
\begin{table}[h!]
  \caption{
Comparison of the ANN predicted results with results from the current best upper bounds and from other estimation methods.} 
  \label{tab:results}
  \begin{threeparttable}[b]
  \begin{center}
    \resizebox{\linewidth}{!}{
    \begin{tabular}{c|c|c|c|c} 
      Observable & Upper Bound &  Upper Bound & Estimation\tnote{a} & ANN \\
      & $N_{\rm max}=10$ & $N_{\rm max}=18$ & $N_{\rm max} \leq 10$ & $N_{\rm max} \leq 10$ \\
      \hline
      gs energy ($MeV$)  & -31.688 & -31.977 & -31.892 & -32.024\\ 
       \hline
      gs rms radius ($fm$) & -- & -- & 2.40 & 2.49 \\
      \hline
   \end{tabular}
    }
    \begin{tablenotes}
         \item [a] The Extrap B method \cite{ExtrapolationMethods:2009} for the gs energy and the crossover-point method \cite{BOGNER:200821} for\\ the gs point proton rms radius
         \end{tablenotes}
  \end{center}
  \end{threeparttable}
\end{table}
\captionsetup{font={footnotesize,rm},justification=centering,labelsep=period}%

It is clearly seen from Figures~\ref{fig:Li6_gs_energy_cp} and~\ref{fig:Li6_gs_radius_cp} above that the ANN method results are consistent with the NCSM calculation results using the Daejeon16 NN interaction at $N_{\rm max} = 12, 14, 16,$ and $18$. Table~\ref{tab:results} also shows that ANN's results are consistent with the best available upper bound in the case of the gs energy. The ANN's prediction for the converged rms radius is slightly larger than the result from the \textit{crossover-point} method and more consistent with the trends visible in Figure~\ref{fig:Li6_gs_radius_cp} at the higher $N_{\rm max}$ values.
To measure the performance of ANNs, MSE for the training subsets up through $N_{\rm max} = 10$, as well as on the second test set for data at $N_{\rm max} = 12, 14, 16,$ and $18$, are provided in Table~\ref{tab:MSE}.

\captionsetup{font={footnotesize,sc},justification=centering,labelsep=period}%
\begin{table}[h!]
  \caption{The MSE performance function values on the training and testing data sets and on the $N_{\rm max} = 12, 14, 16,$ and $18$ data set.}
  \label{tab:MSE}
  \begin{center}
    \resizebox{\linewidth}{!}{
    \begin{tabular}{c|c|c|c|c} 
      Data Set & Whole Set & Training Set & Testing Set$_1$ & Testing Set$_2$ \\
                    & $N_{\rm max}\leq10$ & $N_{\rm max}\leq10$ & $N_{\rm max}\leq10$ & $N_{\rm max}=12-18$ \\
      \hline
      gs energy ($MeV$)  & $4.86\times10^{-4}$ & $5.04\times10^{-4}$ & $3.80\times10^{-4}$ & 0.0072\\
       \hline
      gs rms radius ($fm$) & $7.88\times10^{-7}$ & $4.49\times10^{-7}$ & $2.74\times10^{-6}$ & $9.24\times10^{-7}$ \\
    \end{tabular}
    }
  \end{center}
\end{table}
\captionsetup{font={footnotesize,rm},justification=centering,labelsep=period}%

The small values of the performance function in Table~\ref{tab:MSE} above indicate that ANNs with good generalizations were found to predict the results.

\section{Conclusion and Future Work}\label{sec:Conclusion}



Feed-forward ANNs were used to predict the properties of the $^6\mathrm{Li}$ nucleus such as the gs energy and the gs point proton rms radius. The advantage of the ANN method is that it does not need any mathematical relationship between input and output data.
The architecture of ANNs consisted of three layers: two neurons in the input layer, eight neurons in the hidden layer and one neuron in the output layer. An ANN was designed for each output. 

The data set from the ab initio NCSM calculations using the Daejeon16 NN interaction and basis spaces up through $N_{\rm max} = 10$ was divided into two subsets: 85\% for the training set and 15\% for the testing set. Bayesian regularization was used for training and doesn't require a validation set.

The designed ANNs were sufficient to produce results for these two very different observables in $^6\mathrm{Li}$ from the ab initio NCSM. The gs energy and the gs point proton rms radius showed good convergence patterns and satisfy the theoretical physics condition, independence of basis space parameters in the limit of extremely large matrices. Comparisons of the results from ANNs with established methods of estimating the results in the infinite matrix limit are also provided.
By these measures, ANNs are seen to be successful for predicting the results of ultra-large basis spaces, spaces too large for direct many-body calculations.

As future work, more $\mathrm{Li}$ isotopes such as $^7\mathrm{Li}$, $^8\mathrm{Li}$ and $^9\mathrm{Li}$ will be investigated using the ANN method and the results will be compared with results from improved extrapolation methods currently under development.



\section*{Acknowledgment}

This work was supported by the Department of Energy under Grant Nos. DE-FG02-87ER40371 
and DESC000018223 (SciDAC-4/NUCLEI). The work of A.M.S. was supported by the Russian Science Foundation under Project No. 16-12-10048. Computational resources were provided by the National Energy Research Scientific Computing Center (NERSC), which is supported by the Office of Science of the U.S. DOE under Contract No. DE-AC02-05CH11231.
Personnel time for this project was also supported by Iowa State University.



%
%
%

\bibliographystyle{IEEEtran}
\bibliography{bibtemplate_ANN_Nuclear_Physics_2018_short} 

\end{document}